\documentclass[preprint2]{aastex}
\newcommand{\be}{\begin{equation}}
\newcommand{\ee}{\end{equation}}
\newcommand{\bea}{\begin{eqnarray}}
\newcommand{\eea}{\end{eqnarray}}
\begin{document}
\title{The Anisotropy of MHD Alfv\'{e}nic Turbulence}
\author{Jungyeon Cho \altaffilmark{1}}
\affil{Department of Astronomy, University of Texas, Austin, TX
78712; cho@astro.as.utexas.edu}

\and

\author{Ethan T. Vishniac}
\affil{Department of Physics and Astronomy, Johns Hopkins University,
Baltimore, MD 21218; ethan@pha.jhu.edu}
\altaffiltext{1}{Department of Physics and Astronomy, Johns Hopkins University,
Baltimore, MD 21218}


\begin{abstract}
We perform direct 3-dimensional numerical simulations for
magnetohydrodynamic (MHD) turbulence in a periodic box
of size $2\pi$ threaded by strong uniform
magnetic fields.
We use a pseudo-spectral code with hyperviscosity and hyperdiffusivity
to solve the incompressible MHD equations.
We analyze the structure of the eddies
as a function of scale.
A straightforward calculation of anisotropy in wavevector space
shows that the anisotropy is scale-{\it independent}.
We discuss why this is {\it not} the true scaling law and how the curvature of
large-scale magnetic fields affects the power spectrum
and leads to the wrong conclusion.
When we correct for this effect, we find
that the anisotropy of eddies
depends on their size: smaller eddies are
more elongated than larger ones along {\it local} magnetic field lines.
The results are consistent with the scaling law
$\tilde{k}_{\parallel} \sim \tilde{k}_{\perp}^{2/3}$ proposed by
Goldreich and Sridhar (1995, 1997).  
Here $\tilde{k}_{\|}$ (and $\tilde{k}_{\perp}$) are wavenumbers measured
relative to the local magnetic field direction.
However, we see some
systematic deviations which may be a sign of limitations to 
the model, or our inability to 
fully resolve the inertial range of turbulence in our 
simulations.
\end{abstract}
\keywords{ISM:general-MHD-turbulence}

\section{Introduction}
Many astrophysical plasmas, including the interstellar medium 
and the solar wind,
often show magnetic fields whose energy density is 
greater than or equal to the local
kinetic energy density. In these plasmas the magnetic fields
play a dominant dynamical role, mediated
by magnetohydrodynamic (MHD) waves.
In the incompressible limit, there are only two types of linear modes:
shear Alfv\'{e}n waves and pseudo Alfv\'{e}n waves.
While these two modes have different polarization directions,
they have the same dispersion relation and propagate along
the magnetic field lines at the Alfv\'{e}n speed.
Therefore, the nonlinear interactions of wave packets
moving along the magnetic field lines at the Alfv\'{e}n speed
determine the dynamics of incompressible magnetized plasmas with a 
strong background field.
In this paper, we study the anisotropy of
the MHD turbulence in this regime. We will refer to
this turbulence as incompressible Alfv\'{e}nic turbulence.

Nonlinear processes and the corresponding energy spectrum of 
incompressible Alfv\'{e}nic turbulence are still among
the most controversial problems in MHD.
Since the pioneering  works of Iroshnikov (1963) and Kraichnan (1965),
the Iroshnikov-Kraichnan (IK) theory has been widely accepted as a
model for incompressible, highly conducting MHD turbulence.
The IK theory predicts
 $E_M(k) \sim E_K(k) \sim k^{-3/2}$ from a Kolmogorov-like dimensional
analysis.  Here, $E_M(k)$ and $E_K(k)$
are the magnetic and kinetic energy spectra respectively.
In this framework, two counter-traveling eddies
(i.e. Alfv\'{e}n wave packets)
interact and transfer energy to smaller spatial scales only when they 
collide, as they move in opposite directions along 
the magnetic field lines.
Since the duration of such a collision is shorter than the 
conventional eddy turnover
time by a factor of $t_v(l)/t_A(l)$, this collisional process is 
inefficient and the spectral energy
transfer time as a function of scale $l$ ($=t_{cas}(l)$) increases
by the same factor compared to
the eddy turnover time ($l/v_l$) in ordinary hydrodynamic turbulence.
Here $t_v(l)=l/v_l$ and $t_A(l)=l/V_A$ are
eddy turnover time and Alfv\'{e}n time respectively,
$V_A \equiv B/\sqrt{4\pi \rho}$, and
$B$ is the rms magnetic field strength.  When the 
external field is strong, as assumed in IK theory, this 
quantity is usually set to $B_0$, the strength of
the uniform background field.
If the spectral energy cascade rate
\be
  \epsilon \sim \frac{ v_l^2 }{ t_{cas}(l) } \sim
  \frac{ v_l^3 }{l} \frac{ t_A(l) }{ t_v(l) }
\ee
is scale-independent and $E_M(k) \approx E_K(k)$, then we obtain
the IK energy spectra.

The IK theory assumes an isotropic distribution of energy in 
${\bf k}$-space.
However, many researchers have argued that anisotropy
is an important characteristic in MHD turbulence
(for example, Shebalin et al 1983, Montgomery and Matthaeus 1995).
This anisotropy results from the resonant conditions for 3-wave 
interactions (or 4-wave
interactions, when 3-wave interactions are null).
The resonant conditions for the 3-wave interactions are
\begin{eqnarray}
  {\bf k}_1 + {\bf k}_2 & = & {\bf k}_3, \\
  \omega_1 +  \omega_2 & = & \omega_3,
\end{eqnarray}
where ${\bf k}$'s are wavevectors and $\omega$'s are wave frequencies.
The first condition can be viewed as momentum conservation
and the second as energy conservation.
Alfv\'{e}n waves satisfy the dispersion relation: $\omega = V_A |k_{\|}|$,
where $k_{\|}$ is the component of wavevector parallel to the background
magnetic
field.
Since only opposite-traveling wave packets interact, ${\bf k}_1$ and
${\bf k}_2$ must have opposite signs.
Then from equations (2) and (3),
either $k_{\|,1}$ or $k_{\|,2}$ must be equal to 0.
That is, zero frequency modes are essential
for energy transfer. If $k_{\|,2}=0$, we have
\begin{eqnarray}
  k_{\|,1} &=& k_{\|,3},\\
  k_{\|,2} &=& 0
\end{eqnarray}
(Shebalin et al 1983).
Therefore, in the wavevector space, 
3-wave interactions make energy cascade
in directions perpendicular to the mean magnetic field.
Since the energy cascade is strictly perpendicular to the 
mean magnetic field, the actives modes in wavevector space have a 
slab-like geometry with a constant width.  The implication is
that the nonlinear cascade of energy works against isotropy in
${\bf k}$ space.  Furthermore, it is important to note that 
equations (2) and (3) are true only when
wave amplitudes are constant.  
In reality, nonlinear interactions provide a natural broadening
mechanism, following the uncertainty relation,
$\Delta t \cdot \Delta \omega \sim 1$.
In particular, if a wave has a frequency less than or comparable to the
nonlinear interaction rate, it is effectively a zero frequency mode.

Goldreich and Sridhar (1995, 1997)
showed that in the strong incompressible shear 
Alfv\'{e}nic turbulence regime (i.e. $\tau_{NL}^{-1} \sim \omega$),
these arguments lead
to a new scaling law with a scale-dependent anisotropy.
In this model smaller eddies are more elongated.
Their arguments are based on the assumption of
a {\it critically balanced} cascade, $k_{\parallel}V_A \sim k_{\perp}v_l$,
where $k_{\perp}$ and $k_{\|}$ are wave numbers perpendicular
and parallel to the external dc field\footnote{Later in this paper, we
will use $\tilde{k}_{\perp}$ and $\tilde{k}_{\|}$, 
instead of $k_{\perp}$ and $k_{\|}$, to represent 
their scaling relation in a slightly different, 
yet we believe equivalent, viewpoint. 
Here, $\tilde{k}_{\perp}$ and $ \tilde{k}_{\|}$
are wavenumbers with respect to the direction of the local magnetic field, 
not the external field.}.  
The argument given
above for 3-wave interactions makes it clear that $k_{\perp}$
will tend to increase until it becomes important in the
plasma dynamics.  The assumption of strong nonlinearity 
implies that wave packets lose their identity after they
travel one wavelength along the field lines. Consequently
the eddy turnover time
($(k_{\perp}v_l)^{-1}$) is actually the same as Alfv\'{e}nic time
($(k_{\parallel}V_A)^{-1}$).
In this model, the cascade time, $t_{cas}(l)$ can be determined 
without ambiguity:
$t_{cas} \approx (k_{\perp}v_l)^{-1} \approx (k_{\parallel}V_A)^{-1}$.
Since the cascade time is comparable to the period of the Alfv\'{e}n wave,
the 3-wave resonant condition can be violated according to the 
uncertainty relation 
$\Delta \omega \cdot \Delta t \sim V_A k_{\|} \cdot t_{cas}
\sim 1$.
The quantity $k_{\|}$ is the width of the active region in
wavevector space.
{}Finally the assumption of a scale-independent cascade rate
$\epsilon \sim v_l^2/t_{cas}(l)\sim E_{waves}V_A/L$ gives
\bea
  k_{\parallel} \sim k_{\perp}^{2/3} L^{-1/3}
\left({E_{waves}\over V_A^2}\right)^{1/3}, \\
  v_l \sim V_A(k_{\perp}L)^{-1/3}
\left({E_{waves}\over V_A^2}\right)^{1/3},
\eea
where $E_{waves}$ is the wave energy per mass.  These
formulae assume that all scales, from $L$ on down, are
within the inertial range of MHD turbulence.
Here the typical $k_{\|}$ should be interpreted as the
size of the range of parallel wavevectors, corresponding to a given
$k_{\perp}$, that contain significant energy.

Matthaeus et al. (1998) recently tested this model
numerically, and showed that the anisotropy of low frequency
MHD turbulence scales linearly with the ratio of
perturbed and total magnetic field strength ($b/B$),
a result which seems inconsistent with Goldreich and
Sridhar's model.  To explain this scaling relation, they suggested
that the region of Fourier space where the energy transfer
takes place actively is given by
\be
|{\bf k}\cdot\frac{{\bf B}_0}{\sqrt{4\pi\rho}} |< \frac{1}{ \tau_{NL} },
\ee
where $\tau_{NL}$ is
the eddy turnover time of the energy containing length $L$.
Consequently, ``the region of the wave number
space where spectral transfer is most vigorous'' has a slab-like 
geometry with a constant width proportional to $1/(\tau_{NL}B_0)$.

All these theories (except the IK theory) share a common prediction for
anisotropy: anisotropy should be more pronounced on smaller scales.
Oughton et al. (1994) and Ghosh and Goldstein (1997) already reported
this scale-dependent anisotropy through numerical simulations.
The former extended Shebalin et al. (1983)'s 2-D calculations
to 3-D cases. To measure anisotropy, 
they introduced the Shebalin angles, $\theta_Q$, defined by
$
 \tan^2{\theta_Q}=( \sum k_{\perp}^2 |{\bf Q}({\bf k},t)|^2 )/
                       ( \sum    k_{\|}^2 |{\bf Q}({\bf k},t)|^2  ),
$
where ${\bf Q}$ is vector potential {\bf A}, magnetic field {\bf B}, 
or current {\bf J}, etc.
Greater $\theta_Q$ means greater anisotropy.
They found that $\theta_A < \theta_B < \theta_J$.
If the energy spectrum of {\bf B} scales as $E_M(k)\propto k^{-s}$,
then the spectra of vector potential and current scale as 
$E_A(k)\propto k^{-s-2}$
and $E_J(k)\propto k^{-s+2}$, respectively.
The spectra of vector potential has the steepest slope among the three spectra.
This means that 
the vector potential is least strongly dependent on small scales (and
the current is most strongly dependent on small scales).
Therefore they concluded that anisotropies are more pronounced 
at smaller scales\footnote{
The width of the active region in Fourier space, $k_{\|}$, is a function
of $k_{\perp}$. If 
$$ \left[\frac{ k_{\parallel} }{ k_{\perp} } \right]_{\mbox{small $k_{\perp}$}} 
  >\left[\frac{ k_{\parallel} }{ k_{\perp} } \right]_{\mbox{large $k_{\perp}$}},$$
then we will have the ordering of Shebalin angles as observed 
by Shebalin et al.\ and Oughton et al.
However, if $k_{\|} \propto k_{\perp}$, then we do not expect any ordering
among the angles. In \S3, we will find that,
no matter what the true functional form $k_{\|} = k_{\|}(k_{\perp})$ is,
Fourier transformation smooths out the {\it true} relation and 
leads to a {\it fake} linear relationship between $k_{\|}$ and $k_{\perp}$.
This suggests that their results are in contradiction to our discussion in \S3.
However, the apparent linear relationship between
$k_{\|}$ and $k_{\perp}$ in \S3 is not perfectly linear.
Instead, we expect $k_{\parallel}=c_1k_{\perp}+c_2$, where $c_1$ 
is a decreasing function of $B_0$ 
and $c_2$ depends on the $k_{\|}$ of forcing terms 
(or initial values of $k_{\|}$ for decaying turbulence).
The presence of $c_2$ does not seem to be important 
in our Fig. 3 in \S3.
However, it does affect the calculation of Shebalin angles.
That is, because of $c_2$, 
the ratio $k_{\|}/k_{\perp}=c_1 +  c_2/k_{\perp}$ becomes a function of
$k_{\perp}$. 
At the largest energy containing eddy scale, $k_{\perp} \sim c_2$ and
hence $k_{\|}/k_{\perp}\sim c_1 +O(1)\sim O(1)$.
But, at small scales, the ratio can be much smaller than
unity. Therefore, because of $c_2$, we can obtain
a scale-dependent anisotropy:
$(k_{\|}/k_{\perp})$ at small $k_{\perp}$ is greater than that at large 
$k_{\perp}$.
Note that this is a result of the initial conditions, rather than a true
scaling relation.
This will lead the ordering of the angles as
observed by Shebalin et al.\ and Oughton et al..
This interpretation is qualitatively consistant with their results.
For example. the ratio $k_{\|}/k_{\perp}=c_1 +  c_2/k_{\perp}^{peak}$, 
therefore $(\tan{\theta_Q})^{-1}$,
approaches to a constant value $c_2/k_{\perp}^{peak}$ 
as $B_0$ becomes strong.
Here $k_{\perp}^{peak}$ is the 
wavenumber of the peak of the energy spectrum.
It is also possible to explain the increased anisotropy 
at high magnetic Reynolds
numbers. 
If the magnetic Reynolds number increases, then $k_{\perp}^{peak}$ increases
 and,
therefore, the ratio decreases. }.
They also found a similar ordering for velocity field (and vorticity, etc).
On the other hand, Ghosh and Goldstein (1997) calculated the Shebalin angles
as a function of wavenumber bin.
They found that the Hall-MHD\footnote{Hall MHD includes 
the Hall term, which is important at ion-cyclotron
scales. In this paper, we consider only the standard MHD equations.} 
simulations show
increased anisotropy at small scales (i.e. greater Shebalin angles
at smaller scales). However their standard MHD simulations
do not show increased anisotropies at small scales.
We refrain from comparing their work with ours because they used different 
physics (the Hall effect) and their
simulations are $2\slantfrac{1}{2}$ dimensional. 
The simulations given in this paper are 3-D standard MHD simulations.
We note that none of the previous papers quantitatively compared their
results with particular theories of anisotropy.

In this paper, we examine the scaling law for Alfv\'{e}nic MHD
turbulence numerically and resolve the controversy concerning 
the anisotropic structure of the turbulence.
Our results are consistent with Goldreich and Sridhar's
model. We describe our numerical methods in \S 2.
In \S 3, we describe our results for anisotropy in wavevector space.
In this section, we demonstrate that none of the scaling laws mentioned
above agrees with our data, and explain why a straightforward
evaluation of the distribution of spectral power does not
correspond to a physically meaningful set of scaling laws.
We attribute this to the systematic effects of large scale
curvature of the magnetic fields. 
In \S 4, we determine the shape of individual eddies, avoiding the
systematic error described above.
We compare our results to Goldreich and Sridhar's model
and give our conclusions in \S 5.

\section{Numerical Methods and General Results}
We have employed a pseudospectral code to solve the 
incompressible MHD equations in a periodic box of size $2\pi$:
\begin{equation}
\frac{\partial {\bf V} }{\partial t} = (\nabla \times {\bf V}) \times {\bf V}
      -V_{A0}^2 (\nabla \times {\bf B})
        \times {\bf B} + \nu \nabla^{2} {\bf V} + {\bf f} + \nabla P' ,
        \label{veq}
\end{equation}
\begin{equation}
\frac{\partial {\bf B}}{\partial t}= {\bf B} \cdot \nabla {\bf V}
     - {\bf V} \cdot \nabla {\bf B} + \eta \nabla^{2} {\bf B} ,
     \label{beq}
\end{equation}
where $\bf{f}$ is random driving force
and $P'\equiv P + {\bf V}\cdot {\bf V}/2$. 
Other variables have their usual meaning.
$V_{A0}=B_0/(4\pi\rho)^{1/2}$ is
the Alfv\'{e}n velocity of
the background field, which is set to be order unity in our simulations.
In pseudo spectral methods, we calculate the temporal evolution of
the equations (\ref{veq}) and (\ref{beq}) in Fourier space.
To obtain the Fourier components of nonlinear terms, we first calculate
them in real space, and transform back into Fourier space.
We use 21 forcing components with $2\leq k \leq \sqrt{12}$.
Each forcing component has correlation time of one.
The peak of energy injection occurs at $k\approx 2.5 $.
The amplitudes of the forcing components are tuned to ensure $V \approx 1$.
Since we expect the perturbed magnetic field strength $b$ to be
comparable to or less than $V$ (i.e. $b \leq 1$),
the resulting turbulence can be regarded as strong MHD turbulence
(i.e. $V \sim B_0 \sim B$).
The average helicity in these simulations is not zero. However, 
our results are insensitive to the value of helicity, possibly
because the box size is only slightly bigger than the energy
injection scale.
We use an appropriate projection operator to calculate 
$\nabla P'$ term in
{}Fourier space and also to enforce divergence-free condition
($\nabla \cdot {\bf V} =\nabla \cdot {\bf B}= 0$).
We use up to $256^3$ collocation points.
We use integration factor technique for kinetic and magnetic dissipation terms
and leap-frog method for nonlinear terms.
We eliminate $2\Delta t$ oscillation of the leap-frog method by using
an appropriate average.
At $t=0$, magnetic field has only uniform component and velocity has a support
between $2\leq k \leq 4$ in wavevector space.

We use hyperviscosity and hyperdiffusivity for dissipation terms
to maximize the inertial range. 
The only exception is Run 256P-$B_0$1, where we use physical viscosity and diffusivity.
The power of hyperviscosity
is set to 8, so that the dissipation term in the above equation
is replaced with
\be
 -\nu_8 (\nabla^2)^8 {\bf V},
\ee
where we set the value of $\nu_8$ from the condition $\nu_h (N/2)^{2h} \Delta t
\approx 0.5$ (see Borue and Orszag 1996).
Here $\Delta t$ is the time step and $N$ is the number
of grids in each direction.
We use exactly same expression for the magnetic dissipation term.
We list parameters used for the simulations in Table 1.
We use the notation 256X-Y, where X = H or P refers to hyper- or physical viscosity;
Y = $B_0$1 or $B_0$0.5 refers to the strength of the external magnetic fields.

{}Fig. 1 shows the evolution of $V^2$ and $B^2$ from 
Runs 256H-$B_0$1 and REF2.
Results of 256H-$B_0$1 are plotted as solid lines, and results of REF2 are plotted
as dotted lines. Both runs have similar results for the overlapped time, 
which means that the values of $V^2$ and $B^2$ are not sensitive to the spatial resolution.
The magnetic energy density grows fast at the beginning of the simulations,
as a result
of the stretching of external field lines by turbulent motion.
Since the external field is strong, magnetic forces
soon become strong enough to balance the stretching effect.
This balance happens at $t \sim 1$ and, after a transient stage ($1\leq t\leq 5$),
the turbulence reaches the saturation stage ($t\geq 5$).
In this case the saturation stage begins quite early, 
since the external uniform magnetic field 
is strong and the magnetic diffusivity is effectively 0. 
In general, when the external field is weaker and/or magnetic 
diffusivity is larger, the saturation stage occurs later.
At the saturation stage, $B^2 \sim 1.45$ and $V^2 \sim 0.6$.
Since $b^2 = B^2 - B_0^2 (\equiv 1) \approx 0.45$, 
there is a rough energy equipartition between $b$ and $V$.
Note that, since $V \sim B$, the condition for strong MHD turbulence is 
met.

\section{Fourier Analysis: A False Scaling Law?}
\subsection{Fourier Analysis}

In this subsection, we investigate the spectral 
structure of Alfv\'{e}nic turbulence.
It is important to note that we lose some
information
about individual eddies when we perform a global transformation,
such as the Fourier transformation.
In particular, the scaling law given in this subsection may $not$ be 
true for individual eddies (See \S 3.2 for details).

We plot the 1-dimensional energy spectra $E_K(k)$ and $E_M(k)$ in Fig. 2.
Both spectra peak at the same $k$, 
which is also the scale of the energy injection.
This reflects that there is a rough energy equipartition between $b$ and $V$ at
the largest energy containing scale.
In general, $E_K(k)$ always peaks at the energy injection scale.
However, $E_M(k)$ peaks at a larger k than $E_K(k)$ when
the external field is weak.
The inertial range exhibits of two different scaling ranges: 
a small $k$ range where the slopes are steep
and a large $k$ range where the slopes are mild. 
We believe that the former reflects a true inertial range,
whereas the latter is a result of the $bottleneck$ effect.
The $1/k$ bottleneck effect is a common feature in
numerical hydrodynamic simulations with hyperviscosity
(see Borue and Orzag 1996, 1995).
Interestingly enough, the slope of the bottleneck is actually steeper than
$-1$.
This might mean that the bottleneck effect is less serious in the
MHD case.
The kinetic  and magnetic energy spectra have slightly different
powers in the inertial range.
Although the power indexes of both spectra are close to 
$-5/3$ in the (true) inertial range,
the data are also compatible with IK theory, where the power index is $-3/2$.

We plot normalized 3-dimensional energy spectra of Run 256H-$B_0$1 in Fig. 3.
In the figure, we plot contours of same
\be
  E_3(k_{\perp}, k_{\parallel}) / E_3(k_{\perp}, 0)
\ee
in the $ (k_{\perp},k_{\parallel})$ plane.
As the energy cascades
in the directions perpendicular to the mean field, this ratio drops
as we move away from the $k_{\perp}$ axis.
The figure implies most of the energy is confined in a region around the
$k_{\perp}$ axis.
The thickness of the region depends on the strength 
of the external magnetic fields (Fig. 4).
We use the contours of $ E_3(k_{\perp}, k_{\parallel}) / E_3(k_{\perp}, 0)=0.5$
to measure the thickness. The angle $\Theta_{0.5}$ is the angle
formed by the contours and the horizontal axis.
We can see that $\tan \Theta_{0.5}$ ($= k_{\|}/k_{\perp}$)
is proportional to $b/B_0$.
This result confirms the scaling relation found by Matthaeus et al. (1998).
Both velocity and magnetic fields have very similar structures.
{}From the fact that $k_{\|} < k_{\perp}$ in the active region (see Fig. 3), 
one can conclude that eddies do have anisotropic structure: eddies
are stretched along the direction of the mean field. 
Apparently, Fig. 3 suggests that $k_{\|} \propto k_{\perp}$ and, hence, that
anisotropy is scale-independent.
No theory mentioned in \S1 agrees with our result.
However, it is not clear from the figure whether or not
the true anisotropy is a function of scale:
although the contours show a linear relationship between $k_{\|}$ and $k_{\perp}$,
the relationship doesn't mean that 
all individual eddies have the same major axis to minor axis ratio.
As explained in next subsection, the
$rotation$ of eddies by large-scale waves in the magnetic fields
can distort the actual scaling relation
and lead to the linear relationship shown in the figure.

{}Fig. 5 shows $t_{phase}$ as a function of wavenumber $k$. 
This time scale is defined as the average correlation time $\Delta t$  
such that Fourier components at ${\bf k} = k$ 
have a phase shift of $60^o$ with respect to the original phase.
We plot the result for zero frequency modes (i.e. $k_{\parallel}$=0 
modes). We see that $t_{phase} \sim k^{-1}$ (that is, $k_{\perp}^{-1}$).
How can we interpret this result?  If a turbulent structure, 
characterized by a wavenumber $k$, moves a distance $l$
then the phase is shifted by roughly $kl$, even if the eddy
is unaffected by the motion.  This implies that a large
scale velocity $V$ will change the phase at a rate $kV$, so
that $t_{phase}\propto k^{-1}$.  This implies that our 
calculation of $t_{phase}$ is dominated by large scale
motions rather than by the local cascade of energy, as long
as the nonlinear cascade rate is proportional to $k$ to some
power less than one, which is generally the case.  This is
an example of how large-scale fluctuations can complicate attempts
to find physically meaningful scaling relations. 

\subsection{Rotation Effect}
In the previous subsection, we showed that anisotropy appears to be 
scale-independent. 
In this subsection, we will show that this apparent scaling relation
is an artifact caused by the Fourier transformation.
That is,
we will demonstrate that large-scale modes in the  magnetic field can
distort the actual scaling law at smaller scales
when we perform a Fourier transformation.
In this manner, a straightforward evaluation of anisotropy in 
Fourier space is strongly contaminated by the curvature of the large-scale
magnetic fields and does not reflect the actual local anisotropy
when anisotropy is more pronounced at smaller scales.
Consequently, figures 3 and 4 are compatible with
any scaling law that predicts that smaller eddies are more elongated.

{}Fig. 3 implies that eddies have anisotropic shapes: on average,
eddies are stretched along the direction of ${\bf B}_0$.
However, not all eddies are aligned along the large scale field.
The elongation of an eddy is determined by its interaction
with the $local$ magnetic field, not the background
field.
Since the large-scale magnetic field lines wander with respect to 
${\bf B}_0$, all smaller scale eddies have similar angular 
distributions around $ {\bf B}_0$.
We illustrate this effect in Fig. 6.

In this way, we can explain the results of Goldreich and
Sridhar (1995) and Matthaeus et al. (1998) simultaneously.
Suppose that eddies are oriented according to the local 
field lines (Fig. 6a).
Goldreich and Sridhar's result implies smaller eddies are
relatively more elongated.
When we perform a Fourier transformation and measure the ratio of 
$k_{\|}/k_{\perp}$,
what we actually measure is not the ratio of the minor axis 
to the major axis of individual eddies.
Instead, because the direction of the local magnetic field varies
according to location and Fourier transformation is none other than
a (weighted) averaging process, 
we actually measure the ratio averaged over all possible orientation of
the eddies (Fig. 6b).
The Fourier transformation sees that $L_1$ (and $l_1$) is the minor axis
and $L_2$ (and $l_2$) is the major axis of an eddy. 
The ratio of $L_1/L_2$ ($=l_1/l_2$) 
is determined by the degree of
the {\it wandering} of the large scale magnetic field lines with respect to 
${\bf B}_0$ and, therefore, the
ratio is nearly constant for all eddies, regardless of their sizes and shapes.
In fact, the ratio will depend on the tangent of the angle
between ${\bf B}_0$ and ${\bf B}$.
Since $k_{\perp} \propto 1/L_1$ and $k_{\|} \propto 1/L_2$,
the {\it measured} value of
the ratio $ k_{\|}/k_{\perp}$ is nearly scale-independent.
We expect the angle $\theta$ ($\equiv 90^o-\theta_w$) 
between ${\bf B}_0$ and ${\bf B}$
to be
\be
 \tan{\theta} = b/B_0.
\ee
Therefore, we have
\be
 \sin{\theta} = b/B =\cos{\theta_w},
\ee
which is exactly the scaling relation found by Matthaeus et al.

\section{Measuring Eddy Shapes}

In this section, we analyze the shape of eddies in real space.
As explained in the previous section, we assume that elongated eddies are
aligned in the direction of the {\it local} magnetic fields.
If we want to visualize the shape of eddies, we
need to first identify the direction of the {\it local} magnetic fields.
It is important to note that, although we use the term
`local magnetic fields' for simplicity, the fields are different from
${\bf B}({\bf r})$.
Let us consider an eddy of size $l$. 
The `local magnetic field' of the eddy must act as the `large-scale  magnetic
field' for the eddy.
Therefore, the `local magnetic fields' for eddies of size $l$, must be
smoothly varying vector fields whose characteristic length of
variation is $>l$.
Note also that another eddy of size $l^{\prime}$ ($\neq l$) at the same 
location can have a slightly different direction for the 
`local magnetic field.'
The `local magnetic fields' are functions of position (${\bf r}$)
and eddy size ($l$).
Then, how can we define the direction of the {\it local} magnetic fields?
We implement
2 independent numerical algorithms to calculate the
direction of the $local$ magnetic fields.

In the first method, the local magnetic fields are calculated by
\be
  {\bf B}_l  = ( {\bf B}({\bf r}_1) +{\bf B}({\bf r}_2) )/2
\ee
and the second order structure functions for $v$ and $b$ are given by
\bea
  F_2^v (R,z)= < |{\bf V}({\bf r}_1) -{\bf V}({\bf r}_2)|^2 >,\\
  F_2^b (R,z)= < |{\bf b}({\bf r}_1) -{\bf b}({\bf r}_2)|^2 >,
\eea
where $R=|\hat{ {\bf z} }\times ({\bf r}_2-{\bf r}_1 )|,$
$z=\hat{ {\bf z} }\cdot ({\bf r}_2-{\bf r}_1 )$ and 
$\hat{ {\bf z} }= {\bf B}_l/|{\bf B}_l|.$
That is, $R$ and $z$ are coordinates in a cylindrical coordinate system
in which the z-axis is parallel to ${\bf B}_l$ (Fig. 7).
${\bf B}({\bf r})$ and ${\bf b}({\bf r})$ are the total and 
perturbed magnetic fields at a point ${\bf r}$.  As usual, brackets
denotes a spatial average.

In Fig. 8 we plot the second order structure functions in z-R plane.
The horizontal axis (z-axis) is parallel to ${\bf B}_l$.
The contours reflect
the shapes of the eddies. Eddies are  clearly elongated
along the local field lines, and 
smaller eddies are more elongated.
The velocity and magnetic fields show different structure at 
large scales. 
However, their small scale structure is quite similar. 

In Fig. 9, we plot R-intercepts and z-intercepts of the contours.
The R-intercept and z-intercept of a given contour can be regarded as
a measure of $ \tilde{k}_{\perp}$ and $ \tilde{k}_{\|}$ 
for the corresponding eddy scale\footnote{We interpret $ \tilde{k}_{\|}$ 
as the inverse of the major axis of eddies and 
$ \tilde{k}_{\perp}$ as that of the  minor
axis. Since the major axis is assumed to be parallel to the local
magnetic fields, $ \tilde{k}_{\|}$ is the parallel wavenumber
with respect to the {\it local} magnetic field direction. 
On the other hand, in \S3.1, $k_{\|}$ is parallel to 
${\bf B}_0$. Therefore, $ \tilde{k}_{\|}$ and 
$ \tilde{k}_{\perp}$ in this section
have different meanings from $k_{\|}$ and $k_{\perp}$ in \S3.1.}.
{}Fitting the results for velocity fields gives
\begin{equation}
  v: \tilde{k}_{\parallel} \sim \left\{  \begin{array}{ll}
                       \tilde{k}_{\perp}^{0.69}, &  (256H-B_00.5) \\
                       \tilde{k}_{\perp}^{0.70}, &  (256H-B_01) \\
                       \tilde{k}_{\perp}^{0.73}, &  (256P-B_01)
                  \end{array}  \right.
\end{equation}
On the other hand, for the magnetic fields we find
\be
 b: \tilde{k}_{\parallel} \sim \left\{  \begin{array}{ll}
                       \tilde{k}_{\perp}^{0.64}, &  (256H-B_00.5) \\
                       \tilde{k}_{\perp}^{0.50}, &  (256H-B_01) \\
                       \tilde{k}_{\perp}^{0.53}, &  (256P-B_01)
                  \end{array}   \right.
\ee
The velocity fields show good agreement with the scaling relation, 
$ \tilde{k}_{\|} \sim \tilde{k}_{\perp}^{2/3}$,  
proposed by Goldreich and Sridhar (1995).
The power indices are insensitive to the strength of $B_0$ or the form
of viscosity (and diffusivity).
However, the magnetic field shows different scaling behavior.
When the external field is moderately strong (256H-$B_00.5$),
the power index is very close to 2/3.
On the other hand, for stronger external fields (256H-$B_01$ and
256P-$B_01$), the power indices are smaller than 2/3. 
The results are insensitive to the choice of viscosity (and diffusivity).
It is not clear whether the existence of a separate scaling law for the
magnetic field represents a physical effect not included in Goldreich
and Sridhar's model, e.g. the first signs of small scale intermittency
in the magnetic field distribution, or merely the failure of the
numerical models used here to fully resolve the inertial range of
strong MHD turbulence.

In the second method, we employ a completely different approach.
We obtain the local large scale magnetic fields by filtering
out the small scale magnetic fields:
\be
  {\bf B}_{\sigma} ({\bf r}) =
  \sum_{ {\bf r}^{\prime} }
  {\bf B}( {\bf r}^{\prime} ) \phi( |{\bf r}-{\bf r}^{\prime} | ),
\ee
where $\phi(r) \propto \exp(-r^2/\sigma_r^2) $ is a gaussian function.
To determine the shape of small scale eddies (i.e. eddies smaller than
the filter size, $\sim \sigma_r$), we consider the following quantities:
\bea
  {\bf v}_{\sigma}({\bf r}) = {\bf V}({\bf r}) - {\bf V}_{\sigma} ({\bf r}),\\
  {\bf b}_{\sigma}({\bf r}) = {\bf B}({\bf r}) - {\bf B}_{\sigma} ({\bf r}),
\eea
where ${\bf V}_{\sigma}$ and ${\bf B}_{\sigma}$ are filtered fields 
(cf. eq. (20)).
The fields ${\bf v}_{\sigma}({\bf r})$ and ${\bf b}_{\sigma}({\bf r})$
represent small scale fluctuation of velocity and magnetic fields, the shape
of which can be regarded as an adequate approximation of small scale eddies.
{}From these two fields we calculate
the  structure functions
\bea
  F^v(R,z) = | {\bf v}_{\sigma}({\bf r}_2)-{\bf v}_{\sigma}({\bf r}_1) |,\\
  F^b(R,z) = | {\bf b}_{\sigma}({\bf r}_2)-{\bf b}_{\sigma}({\bf r}_1) |,
\eea
where  $R$ and $z$ are similarly
defined as in the first method with
$\hat{ {\bf z} } \parallel {\bf B}_{\sigma} ({\bf r}_1)$ (Fig. 10).

In Fig. 11, we plot the results of the second method.
We can clearly observe the flattening effect:
When the filter size is large, eddies are less anisotropic.
When filter size is small, eddies show highly anisotropic structure.
In the figure, we plot only the results for magnetic fields.
Velocity fields show similar trends.
The thick contours represent $F^b(R,z) = <|{\bf b}_{\sigma}|^2>^{1/2}$.
The values next to the thick contours are $ <|{\bf b}_{\sigma}|^2>^{1/2}$.

This second method is useful for visualizing small scale eddies,
but may not be as useful for quantitative analysis.
The difficulty comes from the fact that there is no well 
defined eddy scale associated with filtered fields 
${\bf v}_{\sigma}({\bf r})$
and ${\bf b}_{\sigma}({\bf r})$.

In this paper, we will not pursue quantitative analysis using the method 2.
Instead, we just wish to point out 
that both methods describe the same scaling law: smaller eddies are
relatively more stretched along the local magnetic field lines.
In particular, the results from the first method are consistent with the relation 
$ \tilde{k}_{\|} \sim \tilde{k}_{\perp}^{2/3}$
proposed by Goldreich and Sridhar (1995).

\section{Discussion and Conclusions}

Here we rederive the scaling law 
$ \tilde{k}_{\parallel} \propto \tilde{k}_{\perp}^{2/3}$
in the framework of 3-wave interactions. 
Except for the use of the uncertainty principle,
the work in this section is independent of Goldreich and Sridhar's 
derivation.
As noted by Goldreich and Sridhar (1995), 3-wave interactions are an 
adequate
proxy for wave-wave interactions of all orders in a strong MHD
turbulence.
As long as we assume the locality of interactions, it is pointless
to distinguish $k_{\|}$ and $k_{\perp}$ from 
$\tilde{k}_{\|}$ and $\tilde{k}_{\perp}$. 
Hence, for simplicity, we use $k_{\|}$ and $k_{\perp}$
instead of $\tilde{k}_{\|}$ and $\tilde{k}_{\perp}$ during the derivation.

Suppose the 3-dimensional energy spectrum is given by
\begin{equation}
  E_3(k_{\perp},k_{\parallel})\equiv |\hat{{\bf V}}({\bf k})|^2
\propto k_{\perp}^{-2\alpha} f,
\end{equation}
where $\hat{{\bf V}}({\bf k})$ is the amplitude of the mode whose wavevector
is ${\bf k}$ and $f(u)$ is a positive, symmetric function of $u$
(cf. equation (7) of Goldreich and Sridhar (1995)) which describes
the power distribution as a function of eddy shape.
If the $width$ (or, $thickness$) of the energy spectrum
in the direction of $k_{\parallel}$
is $k_{\perp}^{\beta}$, then we can write
\be
  E_3(k_{\perp},k_{\parallel})
  \propto k_{\perp}^{-2\alpha} f(k_{\parallel}/k_{\perp}^{\beta}).
\ee
If the $width$ is caused by the uncertainty principle
($\Delta t \cdot \Delta \omega \approx 1$
with $\Delta t \propto t_{cas}(l)$ and $\Delta \omega \propto
k_{\parallel}$), then
\be
  t_{cas}(l) \propto k_{\perp}^{-\beta}.
\ee

Suppose the energy cascade rate
$\epsilon \sim v_l^2/t_{cas}(l)$, where $l = 2\pi/k_{\perp}$,
is scale-independent.
Because $v_l^2 \sim k_{\perp}^{-2\alpha} k_{\perp}^2 k_{\perp}^{\beta}$
($\sim k_{\perp} E(k_{\perp})$, $E(k_{\perp})=$ 1-dimensional spectrum)
and $t_{cas}(l) \sim k_{\perp}^{-\beta}$, we have
\be
  k_{\perp}^{-2\alpha+2+2\beta} = const.
\ee
Therefore,
\begin{equation}
  1-\alpha + \beta = 0.
\label{c1}
\end{equation}

Now, let's pick up a mode at a wavevector ${\bf p}$ and consider nonlinear
interactions with
other wave modes. First, the strength of the interaction with another
mode at ${\bf q}$ is
$ \propto p |\hat{ {\bf V} }({\bf p})| |\hat{ {\bf V} }({\bf q})|$.
(This comes from the $(\nabla \times {\bf V}) \times {\bf V}$ term in equation (9))
Hereafter we assume $p\equiv |{\bf p}|\approx p_{\perp}$.
Second, the number of interactions is $\propto p^2 p^{\beta}$.
This is the number of modes $near$ ${\bf p}$. Here we use locality of 3-wave
interactions.
If the interactions are random, 
the net change of amplitude per unit time will be the
strength of the interaction times the square root of
the number of interactions, or
\be
  |\Delta \hat{ {\bf V} }({\bf p})| \propto
p |\hat{ {\bf V} }({\bf p})| |\hat{ {\bf V} }({\bf q})|\cdot
  (p^2 p^{\beta})^{1/2}.
\ee
Therefore, we have
\be
  t_{cas}  \propto |\hat{ {\bf V} }({\bf p})|/|\Delta \hat{ {\bf V} }({\bf p})|
            \propto p^{-2} p^{\alpha} p^{-\beta/2},
\ee
where we assumed $p\propto q$.
Equating this with $t_{cas}\propto p^{-\beta}$, we can write
\begin{equation}
  \alpha - 2 = -\beta/2.
\label{c2}
\end{equation}
{}From equations (\ref{c1}) and (\ref{c2}), we have
\begin{equation}
  \alpha=5/3, \beta=2/3,
\end{equation}
which is just the result of Goldreich and Sridhar (1995):
\be
  k_{\parallel} \propto k_{\perp}^{2/3}.
\ee
As a consequence, the 3-D energy spectrum becomes
\be
  E_3(k_{\perp},k_{\parallel})\propto k_{\perp}^{-10/3}
    f(k_{\parallel}/k_{\perp}^{2/3}) 
\ee
and the corresponding 1-D spectrum is given by
\be
 E(k) \propto k^{-5/3}.
\ee

In summary, we have shown that the 
anisotropy of Alfv\'{e}nic turbulence depends on
the spatial scales of eddies. In particular, our results confirm
the claim by Goldreich and Sridhar (1995, 1997) that
smaller eddies are relatively more elongated along the direction of the
local magnetic field lines than larger ones.
Quantitative measurements of the anisotropy using the velocity fields
show good agreement with
their proposed scaling law, $k_{\parallel}\sim k_{\perp}^{2/3}$ as long
we interpret these wavenumbers as referring to the
{\it local} magnetic field direction.
However, when the external magnetic field is very strong,
magnetic fields scale somewhat differently, showing a slightly more
rapid increase in anisotropy at smaller scales.

It is important to note that the correct scaling laws depend
on comparing the eddy shape to the {\it local} magnetic
field direction.

As a final note, we wish to stress that our results are not in agreement
with the IK theory. The  IK theory is based on the assumption of isotropy in
wavenumber space, which may be true when the external magnetic field is very
weak or zero.  However, even in these cases, the turbulence is globally isotropic,
but locally very anisotropic.  In this paper, we showed that eddies do show
anisotropy and that the anisotropy is scale-dependent when there is a strong
large scale field (which should apply to very small scales within any MHD
turbulence cascade). On the other hand, our results
are consistent with Goldreich and Sridhar's theory of strong MHD turbulence.
More precisely, if we consider the ratio of hydrodynamic to Alfv\'{e}nic
rates, that is $(k v_k/ \tilde{k}_{\|} V_A)$, we find from equations (18)
and (19) that
\begin{equation}
{k v_k\over \tilde{k}_{\|} V_A}\propto \tilde{k}_{\perp}^{0.3-0.5} v_k,
\end{equation}
where $k\approx \tilde{k}_{\perp}$.
{}From Fig. 2 we see that for the inertial range this implies a ratio
which is either constant or increasing with wavenumber.  A constant
ratio is predicted by Goldreich and Sridhar's model.  The IK model
predicts a slow decline.

\acknowledgements
This work was partially supported by National Computational Science
Alliance under CTS980010N and utilized the NCSA SGI/CRAY Origin2000.

\onecolumn
\clearpage

\begin{deluxetable}{ccccl}
\tablecaption{Parameters}
\tablehead{
\colhead{Run \tablenotemark{a}} & \colhead{$N^3$} & \colhead{$\nu$} & \colhead{$\eta$} &
\colhead{$B_0$}
}
\startdata
REF1 & $144^3$ & $3.20 \times 10^{-28}$  &  $3.20 \times 10^{-28}$  & 0.5
\nl
REF2 & $144^3$ & $3.20 \times 10^{-28}$  &  $3.20 \times 10^{-28}$  & 1
\nl
REF3 & $144^3$ & $3.20 \times 10^{-28}$  &  $3.20 \times 10^{-28}$  & 2
\nl
REF4 & $144^3$ & $3.20 \times 10^{-28}$  &  $3.20 \times 10^{-28}$  & 3
\nl
256H-$B_0$0.5 & $256^3$ & $6.42 \times 10^{-32}$  &  $6.42 \times 10^{-32}$  & 0.5
\nl
256H-$B_0$1 & $256^3$ & $6.42 \times 10^{-32}$  &  $6.42 \times 10^{-32}$  & 1
\nl
256P-$B_0$1 & $256^3$ &  0.001                  &  0.001                   & 1
\nl
\enddata

\tablenotetext{a}{ For $256^3$ grids, we use the notation 256X-Y, 
where X = H or P refers to hyper- or physical viscosity;
Y = $B_0$1 or $B_0$0.5 refers to the strength of the external magnetic fields.}

\end{deluxetable}

\clearpage
\figcaption{Time evolution of $V^2$ and $B^2$ (Solid lines, 256H-$B_0$1;
      Dotted lines, REF2). $B_0= 1$ for both runs.}
\figcaption{256H-$B_0$1. Kinetic energy spectrum ($E_K(k)$)
      and magnetic energy spectrum ($E_M(k)$). For $2\leq k\leq 20$, spectra
      are compatible with $k^{-5/3}$. A 1/k bottleneck effect is observed
      before the dissipation cutoff $k_d \sim 90$. }
\figcaption{256H-$B_0$1. 3D energy spectra normalized by the value on the
$k_{\perp}$ axis.}
\figcaption{$\tan \Theta_{0.5}$ ($=k_{\|}/k_{\perp}$) versus $b/B_0$.
These results are consistent with Matthaeus et al (1998) showing
that anisotropy ($k_{\|}/k_{\perp}$) scales linearly with $b/B_0$.}
\figcaption{256H-$B_0$1. Phase correlation time $t_{phase}$ as a 
function of wavenumber, defined as the average time required for the 
phase to shift by $60^o$. }
\figcaption{Rotation effect. (a) Anisotropic wave packets are moving
      along magnetic field lines. We assume the wave packets are aligned
      with the direction of local magnetic field lines.
      (b) Large scale variations in the magnetic field direction causes
      wave packets to point in different directions as a function of
location.  We expect $L_1/L_2 = l_1/l_2$. In Fourier space,
      $L_1$ (and $l_1$) become $\sim 1/k_{\perp}$ and $L_2$ (and $l_2$) 
becomes
      $\sim 1/k_{\|}$. Consequently, information about 
      eddy shapes is lost when we look at the power spectrum.}
\figcaption{Method 1. We adopt a cylindrical coordinate system in which
      z-axis is parallel to
      ${\bf B}_l  = ( {\bf B}({\bf r}_1) +{\bf B}({\bf r}_2) )/2$.}
\figcaption{256H-$B_0$1. Visualization of eddies using method 1. Note that horizontal
      axis is z-axis (see Fig. 7 for definition). Unit is grid spacing. Smaller eddies are more elongated.}
\figcaption{R-intercept (semi-minor axis; $\sim 1/ \tilde{k}_{\perp}$) versus
      z-intercept (semi-major axis; $\sim 1/ \tilde{k}_{\|}$)
      from Fig. 8. In 256H-$B_0$0.5, both velocity and magnetic fields follow
      the relation $ \tilde{k}_{\|} \sim \tilde{k}_{\perp}^{2/3}$.
      In 256H-$B_0$1 and 256P-$B_0$1, velocity fields follow the same scaling relation.
      However, magnetic fields scale slightly differently.}
\figcaption{Method 2. We adopt a cylindrical coordinate system in which the
      z-axis is parallel to large-scale fields ${\bf B}_{\sigma}$.}
\figcaption{256H-$B_0$1. Visualization of eddies using method 2. Note that the horizontal
      axis is the z-axis (see Fig. 10 for definition). 
      Unit is grid spacing. Only magnetic fields are shown.
      Smaller eddies are more flattened.}
\clearpage
\begin{figure}
\plotone{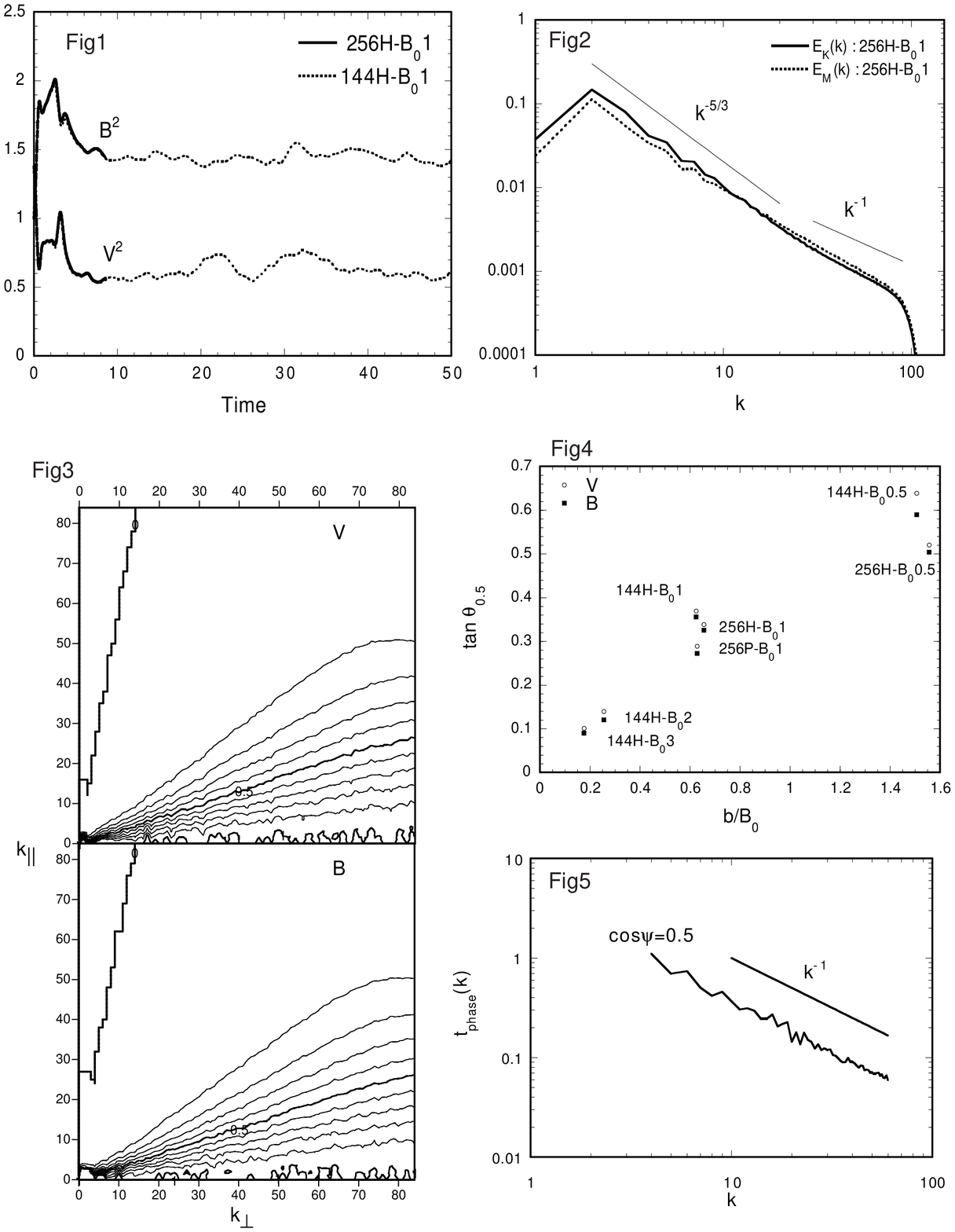}
\end{figure}
\clearpage
\begin{figure}
\plotone{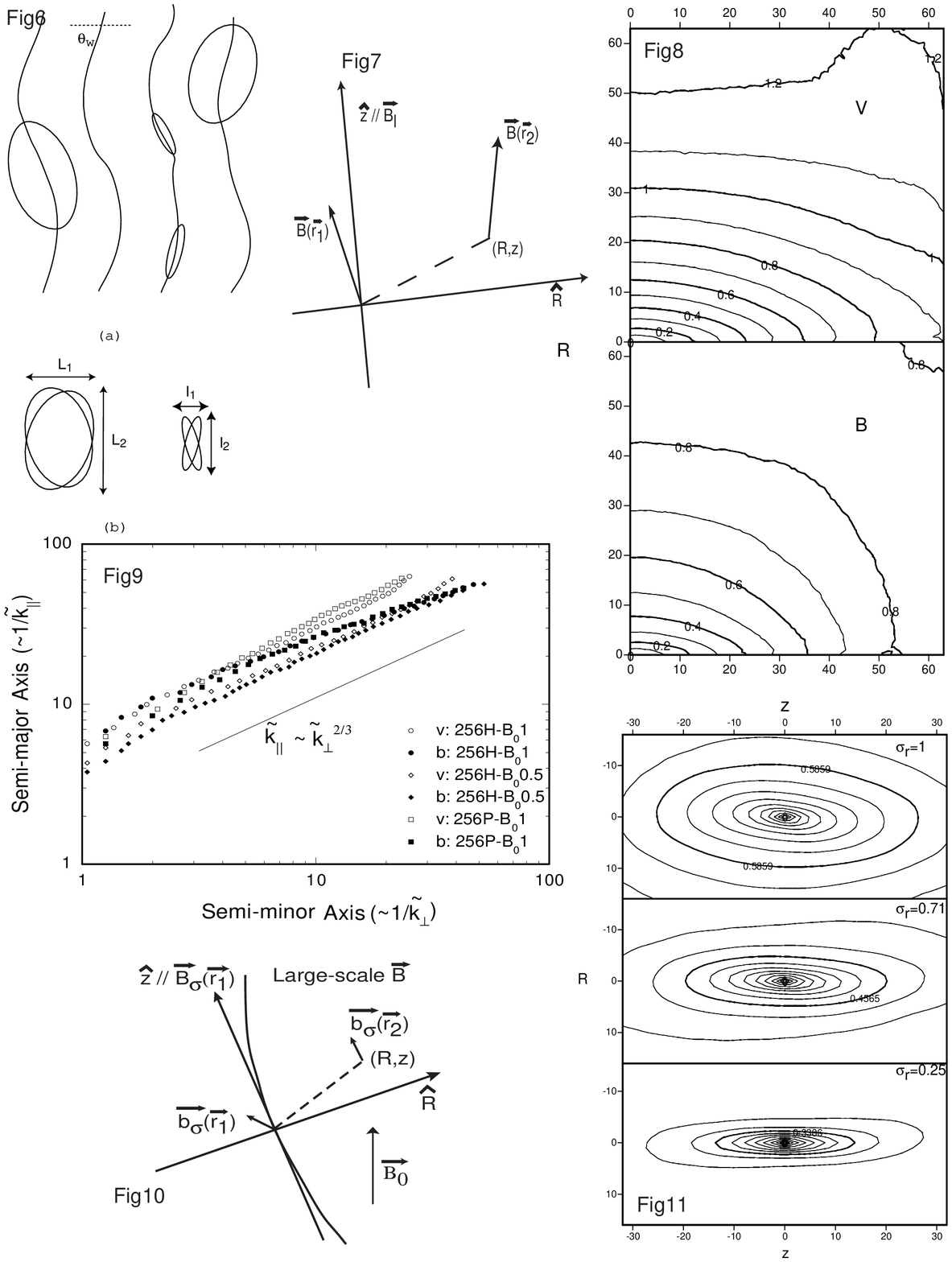}
\end{figure}
      
\end{document}